\documentclass[epj,nopacs]{svjour}
\usepackage[utf8]{inputenc}
\usepackage{amsmath,amssymb,amsfonts,graphicx,cite,multirow,ulem}
\usepackage[retainorgcmds]{IEEEtrantools}
\graphicspath{{graphics/}}
\makeatletter
\ifx\input@path\@undefined
\def\input@path{{graphics/}}
\else
\g@addto@macro\input@path{{graphics/}}
\fi
\makeatother

\newcommand{\hw}{\textsf{Herwig 7}}
\usepackage{color}

\preprint{%
MAN/HEP/2016/09\\
CERN-TH-2016-124\\
IFJPAN-IV-2016-12\\
HERWIG-2016-05\\
MCnet-16-16\\
IPPP/16/48
}

\title{Reweighting Parton Showers}

\author{Johannes Bellm\inst{1} \and Simon Pl\"atzer\inst{1,2}\and Peter
  Richardson\inst{3,1} \and Andrzej Si\'odmok\inst{3,4} \and Stephen Webster\inst{3,1}}

\institute{
IPPP, Department of Physics, Durham University\and
Particle Physics Group, School of Physics and Astronomy, University of Manchester\and
Theory Department, CERN, Geneva\and
The Henryk Niewodniczanski Institute of Nuclear Physics in Cracow, Polish Academy of Sciences}

\date{\today}

\abstract{We report on the possibility of reweighting parton-shower Monte
  Carlo predictions for scale variations in the parton-shower algorithm. The
  method is based on a generalization of the Sudakov veto algorithm. We
  demonstrate the feasibility of this approach using example physical
  distributions.  Implementations are available for both the parton-shower
  modules in the Herwig 7 event generator.\PACS{ {xx.yy.zz}{Xx Yy Zz} } }

\begin{document}

\maketitle


\section{Introduction}

Monte Carlo simulations~\cite{Bahr:2008pv,Bellm:2015jjp,1126-6708-2006-05-026,
  Sjostrand:2014zea,Sjostrand:2007gs,Gleisberg:2008ta} have become essential
tools in both the analysis of data from energy frontier particle physics
experiments and the design of future experiments. The last ten years have seen
a dramatic improvement in the accuracy of these simulations with the
development of techniques to improve the description of high-multiplicity jet
production at leading order\footnote{See Ref.\,\cite{Buckley:2011ms} for a
  recent review of older techniques.}~\cite{Catani:2001cc, Krauss:2002up,
  Lonnblad:2001iq, Schalicke:2005nv, Krauss:2005re, Lavesson:2005xu,
  Mrenna:2003if, Mangano:2002ea, Alwall:2007fs,Hoeche:2009rj,Hamilton:2009ne},
simulations accurate at next-to-leading order, including the description of
the hardest emission at leading order, \cite{Frixione:2002ik,Nason:2004rx},
and more recently multiple jet production at next-to-leading order~(NLO)
\cite{Hoche:2010kg,Hamilton:2010wh,Alioli:2012fc,Frederix:2012ps,Lonnblad:2012ng,
  Platzer:2012bs,Lonnblad:2012ix}. This unprecedented increase in accuracy
means that often the results of modern Monte Carlo event generators are the
main, or even only, way in which theoretical predictions are compared to the
latest results of the LHC experiments. This means that it is vital that
wherever possible we must be able to assess the uncertainty on the predictions
of event generators, as well as the central value of the prediction. These
uncertainties come from a number of different sources:
\begin{itemize}
\item missing higher order corrections in the calculation of the hard matrix elements and
  shower evolution, normally estimated by varying the factorization and
  renormalization scales\cite{Hoche:2012wh,Bellm:2016rhh};
\item uncertainties from the perturbative and\linebreak non-perturbative modelling in
  the event generator, usually estimated by using different tunes of the event
  generator parameters\cite{Richardson:2012bn};
\item uncertainties from the fitting of the parton distribution
  functions~(PDFs), now normally estimated using the recommendations of the
  PDF4LHC working group\cite{Rojo:2015acz,Butterworth:2015oua}.
\end{itemize}
As event generators have become more sophisticated, in the calculation of the
perturbative physics the time taken for the calculation of the hard 
partonic configurations has increased which, together with the time taken for any
simulation of the detector, means that it is often unfeasible to rerun the
event generator for each scale choice, set of parton distribution functions,
and non-perturbative parameters needed to fully assess all the sources of
uncertainty in the Monte Carlo simulation.

In the calculation of the hard process the calculation of the uncertainty from
the variation of the factorization and renormalization scales, together with
the PDF uncertainty, can be more efficiently calculated.  This is achieved by
calculating the effect of changing the scale, or PDF, as a weight with respect
to the central values at the same time as computing the central value. These
weights can then be used to reweight the result of the simulation to obtain
the uncertainties without requiring additional runs of the event
generator. While this does increase the run time of the event generator, over
that of simply performing the calculation for one PDF and scale choice, it is
expected to be much more efficient than fully simulating events for all the
required choices of scales and PDFs.

In contrast, currently the effect of varying the scale in the parton shower, or
any of the perturbative parameters controlling the parton shower, or
non-perturbative parameters for the simulation of the underlying event and
hadronization, can only be calculated by running the full event simulation for
each scale or parameter choice. Given the number of binary
choices which are made in both the parton shower and the various
non-perturbative models it is far from obvious that the variation of the
parameters of these models can be achieved using a reweighting procedure.
 
In this paper we will present a generalization of the Sudakov veto algorithm
used in most modern Monte Carlo event generators to generate the parton
shower. This modification will allow us to calculate the effect of changing
parameters in the parton shower via a reweighting of the central result,
rather than a full resimulation of the events. The first
approach~\cite{Stephens:2007ah} to calculating these weights on an
event-by-event basis required calculating a weight for each variation which
was complicated as this weight was not calculated via the veto algorithm
making it difficult to implement in a full event generator, particularly for
weights involving the PDFs. Performing the changes in the Sudakov veto
algorithm was introduced for final-state radiation in
\cite{Giele:2011cb}. Related work on modifying the Sudakov veto algorithm to
address a number of applications has been presented {\it e.g.} in
\cite{Hoeche:2009xc,Lonnblad:2012hz}, while detailed studies regarding
negative splitting kernels and effects of the infrared cutoff have been
addressed in \cite{Platzer:2011dq}. In the next section we will present the
full details of the algorithm. This is followed by various checks that the
reweighting procedure correctly reproduces the results of resimulating the
events with different parameters and the calculation of the uncertainty for a
number of physical distributions. In this first proof-of-concept paper we will
concentrate on the effect of varying the scale used in the strong coupling and
PDFs, although other changes can be simulated using the same approach. Finally
we present our conclusions and directions for future work.

\section{The Weighted Sudakov Veto Algorithm}

\subsection{Standard Veto Algorithm}

The standard veto algorithm proceeds, given the starting scale $Q$, to
generate the scale of the next emission $q$ and the $d$ additional splitting
variables\footnote{For example if we are considering $1\to2$ splittings $d=2$
  and the splitting variables may be the light-cone momentum fractions of the
  partons and azimuthal angle of the branching.} $x$ according to the
distribution
\begin{IEEEeqnarray}{LCL}
\IEEEeqnarraymulticol{3}{l}{ {\rm d}S_P (\mu,x_\mu|q,x|Q) } \nonumber\\
\qquad = & {\rm d}q\ {\rm d}^dx & \left[ \Delta_P(\mu|Q) \delta(q-\mu) \delta(x-x_\mu)\right. + 
\label{eqn:sud}\\
\qquad & & \left.P(q,x)\theta(Q-q)\theta(q-\mu)\Delta_P(q|Q)\right], \nonumber
\end{IEEEeqnarray}
where $x_\mu$ is a parameter point associated to the cutoff $\mu$, the splitting
kernel is $P(q,x)$ and the Sudakov form factor is
\begin{equation}
\Delta_P(q|Q) = \exp\left(-\int_q^Q {\rm d}k \int {\rm d}^dz\ P(k,z)\right). 
\end{equation}
The distribution $S_P$ is normalized to unity.

The standard veto algorithm proceeds by taking an overestimate of the kernel
$R(q,x)$ such that
\begin{equation}
R(q,x) \geq P(q,x) \ \ \ \ \forall\ \ \  q,x. \label{eqn:over}
\end{equation}
Normally, the overestimate is chosen to be integrable and invertible so $q,x$ 
can easily be generated according to the overestimated distribution
\begin{IEEEeqnarray}{LCL}
\IEEEeqnarraymulticol{3}{l}{ {\rm d}S_R (\mu,x_\mu|q,x|Q) }\nonumber\\
\qquad = & {\rm d}q\ {\rm d}^dx & \left[ \Delta_R(\mu|Q) \delta(q-\mu) \delta(x-x_\mu)\right. + \\
\qquad & & \left. R(q,x)\theta(Q-q)\theta(q-\mu)\Delta_R(q|Q)\right], \nonumber
\end{IEEEeqnarray}
with a Sudakov form factor
\begin{equation}
\Delta_R(q|Q) = \exp\left(-\int_q^Q {\rm d}k \int {\rm d}^dz\ R(k,z)\right). 
\end{equation}
The generation of the splitting scale and variables starting at scale $k=Q$
proceeds as follows:
\begin{enumerate}
\item A trial splitting scale and variables, $q,x$, are generated according to
  $S_R (\mu,x_\mu|q,x|k)$.
\item If the scale $q=\mu$ then there is no emission and the cut-off scale,
  $\mu$, and associated parameter point $x_\mu$ are returned.
\item The trial scale and splitting variables are accepted with probability
\begin{equation}
\frac{P(q,x)}{R(q,x)}, \label{eqn:vetoprob}
\end{equation}
otherwise the process is repeated with $k=q$.
\end{enumerate}
For a proof that this correctly reproduces the distribution in
Eqn.\,\ref{eqn:sud}, see for example{\cite{Buckley:2011ms,Platzer:2011dq}.

\subsection{Weighted Algorithm}

We can generalise the veto algorithm to include weights while simultaneously
relaxing the requirements so that $P$ is not required to be positive and
removing the restriction on $R$, Eqn.\,\ref{eqn:over}. In this case $S_P$ is
still normalized to unity. In order to achieve this we need to introduce an
acceptance probability $\epsilon(q,x|k,y)$ such that
\begin{equation}
0 \le \epsilon(q,x|k,y) < 1.
\end{equation}
In this case we start with a weight $w=1$.  The generation of the splitting
scale and variables together with the calculation of the weight proceeds as
follows:
\begin{subequations}
\label{eqn:weights}
\begin{enumerate}
\item A trial splitting scale and variables, $q,x$, are generated according to
  $S_R (\mu,x_\mu|q,x|k)$.
\item If the scale $q=\mu$ then there is no emission and the cut-off scale,
  $\mu$, and associated parameter point $x_\mu$ are returned with weight $w$.
\item The trial splitting variables $q,x$ are accepted with probability
  $\epsilon(q,x|k,y)$ and the returned weight is 
\begin{equation}
w \times \frac{1}{\epsilon(q,x|k,y)} \times \frac{P(q,x)}{R(q,x)}\label{eq:accept},
\end{equation}
\item Otherwise the weight becomes
\begin{equation} 
w\times \frac{1}{1-\epsilon(q,x|k,y)}\times \left(1-\frac{P(q,x)}{R(q,x)}\right)\label{eq:reject},
\end{equation}
and the algorithm continues with $k=q$.
\end{enumerate}
\end{subequations}
We stress that, in general, the acceptance probability $\epsilon$ can depend
both on the point under consideration for a veto and the previously vetoed point, allowing
the algorithm to be biased to traverse certain sequences more often than
others.  In general the algorithm is not guaranteed to terminate, however this
is not an issue for the applications we are considering.

\subsection{Proof of the Algorithm}

In order to prove that this algorithm gives the correct result we note
that the probability density for the algorithm to traverse a sequence
$(q,x|q_n,x_n|...|q_1,x_1)$ of $n-1$ veto steps to give a result
$q,x$ from an initial condition $Q\equiv q_1,x_Q\equiv x_1$ is
\begin{IEEEeqnarray}{RL}
\IEEEeqnarraymulticol{2}{L}{ {\rm d}S_{R,\epsilon}^{(n)}(\mu,x_\mu;q,x|q_n,x_n|...|q_1,x_1)}\nonumber\\
 = & \left[  \epsilon(q,x|q_n,x_n)\ R(q,x) \theta(q_n-q)\theta(q-\mu)\Delta_R(q|q_1)  \right. \nonumber\\
& + \left. \Delta_R(\mu|q_1)\delta(q-\mu)\delta(x-x_\mu) \right] {\rm d}q\ {\rm d}^dx \nonumber\\
& \times \prod_{i=2}^n R(q_i,x_i)(1-\epsilon(q_i,x_i|q_{i-1},x_{i-1})) \nonumber\\
&\IEEEeqnarraymulticol{1}{c}{\theta(q_{i-1}-q_i)\theta(q_i-\mu)
\ {\rm d}q_i\ {\rm d}^d x_i ,}
\end{IEEEeqnarray}
where we have introduced
an arbitrary parameter point $x_Q$ at the start of the algorithm,
which may be chosen to improve on the efficiency of the
algorithm.

The weight accumulated through such a sequence is
\begin{multline}
w^{(n)}_{P,R,\epsilon}(\mu,x_\mu;q,x|q_n,x_n|...|q_1,x_1) =\\ \prod_{i=2}^n
\frac{1}{1-\epsilon(q_i,x_i|q_{i-1},x_{i-1})}\left(1-\frac{P(q_i,x_i)}{R(q_i,x_i)}\right)\\
\times 
\left\{
\begin{array}{lr}
\frac{1}{\epsilon(q,x|q_n,x_n)} \times \frac{P(q,x)}{R(q,x)} & q > \mu,\\
1 & q = \mu.
\end{array}
\right.
\end{multline}
The density produced by the algorithm is therefore
\begin{eqnarray}
\lefteqn{{\rm d}S_{P,R,\epsilon} (\mu,x_\mu|q,x|q_1,x_1)} \nonumber\\
&=&\sum_{n=1}^\infty \int_{q_2,x_2,...,q_n,x_n}{\rm d}S_{R,\epsilon}^{(n)}(\mu,x_\mu;q,x|q_n,x_n|...|q_1,x_1)
\nonumber\\
&&\ \ \ \ \ \ \ \ \ \ \ \ \ \ \ \ w^{(n)}_{P,R,\epsilon}(\mu,x_\mu;q,x|q_n,x_n|...|q_1,x_1).
\end{eqnarray}
Using
\begin{IEEEeqnarray}{RL}
\IEEEeqnarraymulticol{2}{l}{ {\rm d}S_{R,\epsilon}^{(n)}(\mu,x_\mu;q,x|q_n,x_n|...|q_1,x_1) }\nonumber\\
 \IEEEeqnarraymulticol{2}{c}{ \times \ w^{(n)}_{P,R,\epsilon}(\mu,x_\mu;q,x|q_n,x_n|...|q_1,x_1) }\nonumber\\
 =  & \left[ \Delta_R(\mu|q_1)\delta(q-\mu)\delta(x-x_\mu) \right. \nonumber \\
&  + \left. P(q,x)\theta(q_1-q)\theta(q-\mu)\Delta_R(q|q_1) \right] \ {\rm d}q\ {\rm d}x \nonumber\\
 & \times \prod_{i=2}^n \theta(q_{i-1}-q_i)\theta(q_i-q) \nonumber\\ 
 & \IEEEeqnarraymulticol{1}{c}{\left(R(q_i,x_i)-P(q_i,x_i)\right) {\rm d}q_i\ {\rm d}x_i , }
\end{IEEEeqnarray}
the difference $R(q,x)-P(q,x)$ exponentiates when performing the sum as for
the standard veto algorithm hence
\begin{equation}
{\rm d}S_{P,R,\epsilon} (\mu,x_\mu|q,x|q_1,x_1)={\rm d}S_P(\mu,x_\mu|q,x|q_1),
\end{equation}
{\it i.e.} the correct distribution is produced.

\subsection{Competing Channels}

Often we have to deal with the case of competing processes, for example the
branching of a gluon, $g\to gg$ and $g\to q\bar q$.  This can be correctly
handled using the competition algorithm,\footnote{Sometimes this is referred to
  as ``winner-takes all.''}  {\it i.e.} generating a trial emission for all
the possible processes and then selecting the one with the highest emission scale.

In this case the proof follows the standard one for the competition. We can
continue to use competition to generate the different branchings and still
take the emission with the highest scale but using the weighted Sudakov veto
algorithm the weight is the product of the weights for all the trial
emissions, including those which are rejected.

\subsection{Applications}
There are many potential uses of this weighted Sudakov Veto algorithm. 
One important use which we will not consider here is handling
more complicated kernels, which can be negative, while still generating emissions,
albeit weighted, using standard techniques. 

A second use, which is the main aim of this paper, is that it
gives us a method of performing the parton shower for a default splitting kernel
$P(q,x)$ while at the same time calculating the weights for different choices
of the kernel. Here we will consider the simplest possible choice, {\it i.e.}
changing the scale used in the strong coupling and PDFs, but the method allows
for any variation which can be expressed as a change of the kernel.

In this case we will choose the acceptance probability 
\begin{equation}
\epsilon(q,x|k,y) = \frac{P(q,x)}{R(q,x)}\label{eq:stdveto},
\end{equation}
for the default choice of kernel $P(q,x)$. 
Using this choice the unweighted result reproduces the result of the
standard veto algorithm for the default kernel. While the weighted results will
produce the result for different choices of the kernel. This
choice ensures that in our case the weighted Sudakov veto algorithm will
terminate.  Variations of the splitting kernels can now be introduced by
changing $P\to P'$ in Eqns.~\ref{eq:accept} and \ref{eq:reject}, while
keeping the acceptance probability given in Eqn.~\ref{eq:stdveto}.

\section{Results}

In this section we will show that using the weighted Sudakov veto algorithm
allows us to correctly reproduce the effect of varying the scale of the strong
coupling and PDFs in the parton shower kernels without the need for multiple
runs of an event generator using different scales. We will use this new
approach to study the uncertainty from scale variations for a few example
physical distributions.

We will use \hw\cite{Bellm:2015jjp} for these studies. \hw\ provides the
option of using two different parton-shower algorithms: the default
angular-ordered shower \cite{Gieseke:2003rz} using $1\to2$ branchings together
with global momentum reshuffling to ensure momentum conservation; and a dipole
based approach using local recoils \cite{Platzer:2009jq}. The implementations
of both of these showers use the veto algorithm to generate the emissions, in
the case of the angular-ordered shower using simple overestimate functions and
vetos while for the dipole shower the \textsf{ExSample} library
\cite{Platzer:2011dr} is used to adaptively sample the Sudakov distribution
using the veto algorithm.

\hw\ allows us to compare the results of two physically different
parton-shower algorithms. At the same time we also can check that the weighted
results are correct using two algorithms which differ both in the physical
approach used and in the technical implementation of the veto algorithm in the
program.

Figs.\,\ref{fig:thrustAO} and \ref{fig:thrustDipole} show the differential distribution for \makebox{$1-T$} for
$e^+e^-\to q \bar{q}$ with $\sqrt{s}=91.2\,$GeV for the angular-ordered and dipole showers, respectively.
These results were obtained at the parton level after the parton shower without any corrections
to describe hard radiation.

\begin{figure}
\includegraphics[width=0.5\textwidth]{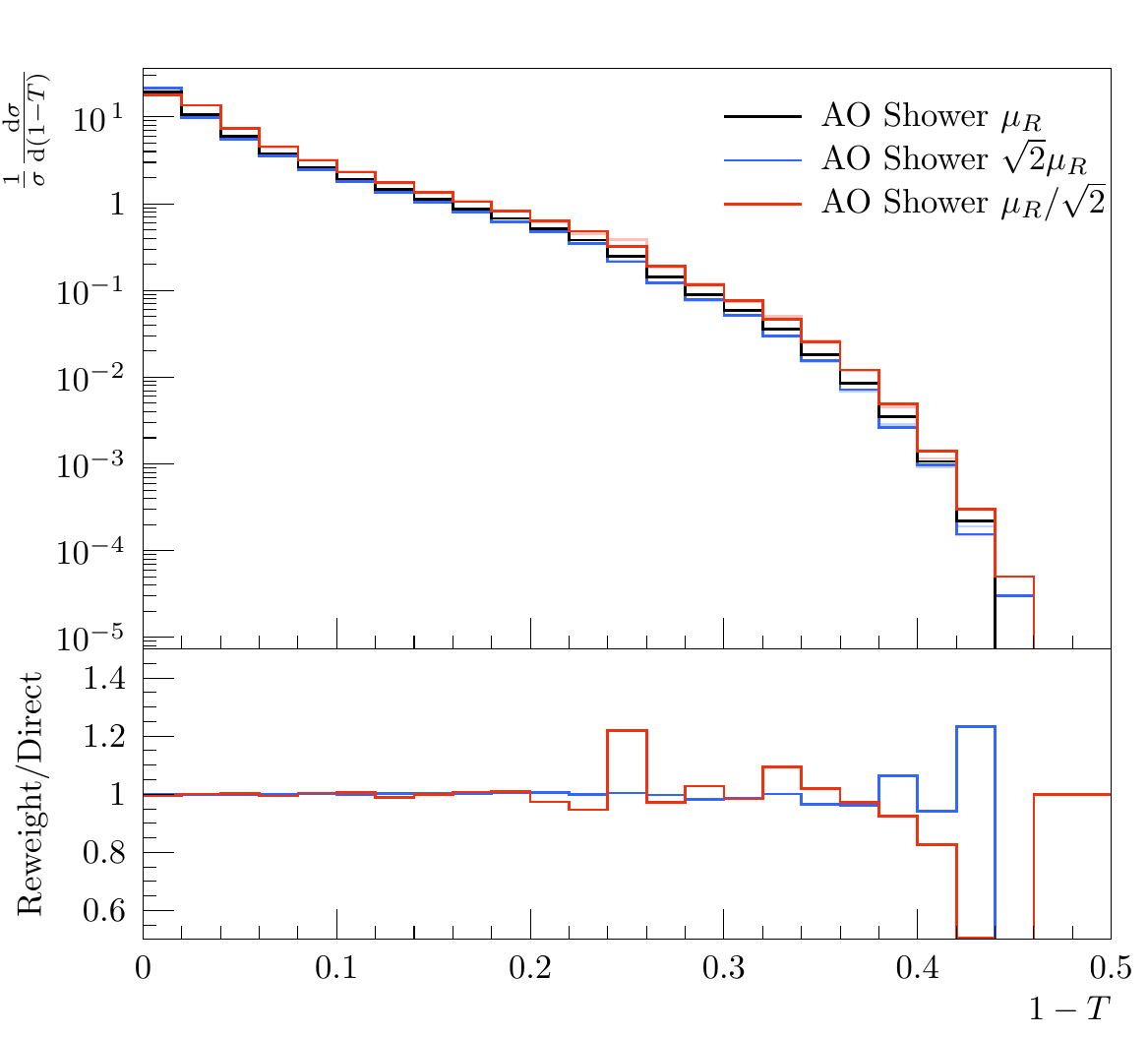}
\caption{Distribution of $1-T$ at the parton level for $e^+e^-\to q \bar{q}$ with $\sqrt{s}=91.2\,$GeV
using the angular-ordered parton shower.
The upper frame shows the effect of varying the scale while the
lower shows the ratio of the up and down scale variations calculated using reweighting
with respect to those obtained running the event generator with the relevant scale.}
\label{fig:thrustAO}
\end{figure}

\begin{figure}
\includegraphics[width=0.5\textwidth]{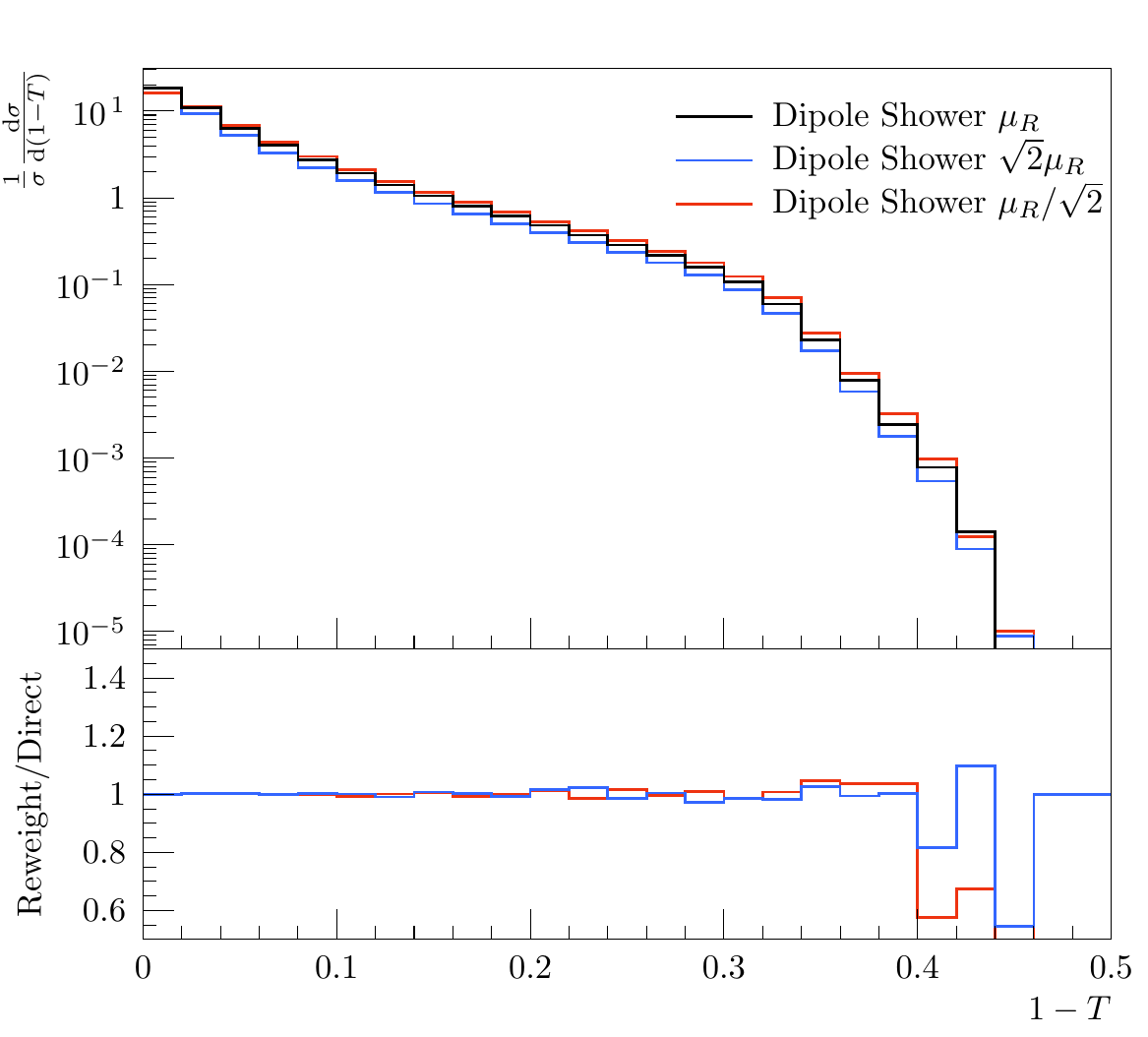}
\caption{Distribution of $1-T$ at the parton level for $e^+e^-\to q \bar{q}$ with $\sqrt{s}=91.2\,$GeV
using the dipole shower.
The upper frame shows the effect of varying the scale while the
lower shows the ratio of the up and down scale variations calculated using reweighting
with respect to those obtained running the event generator with the relevant scale.
}
\label{fig:thrustDipole}
\end{figure}

Figs.\,\ref{fig:higgsAO} and \ref{fig:higgsDipole} show the differential distribution for 
transverse momentum of the Higgs boson for
$gg\to h^0$ with $\sqrt{s}=13\,$TeV for the angular-ordered and dipole showers, respectively.
These results were obtained at the parton level after the parton shower without any corrections
to describe hard radiation.

\begin{table*}
\begin{tabular}{|c|c|ccc|ccc|ccc|}
\hline
Shower & Hadron- & \multicolumn{3}{|c|}{No}  & \multicolumn{6}{|c|}{MPI} \\
\cline{6-11}
& ization & \multicolumn{3}{|c|}{MPI} & \multicolumn{3}{|c|}{Primary} & \multicolumn{3}{|c|}{All} \\
&  \& Decays  & Direct & Reweight & Frac. Diff. & Direct & Reweight & Frac. Diff.& Direct & Reweight & Frac. Diff. \\
\hline
AO & Off & 79.8  & 94.2  & -0.18 & 384.4 & 249.1 & 0.35 & 416.7 & 375.1 & 0.09 \\ 
   & On  & 183.2 & 128.3 &  0.30 & 738.7 & 364.3 & 0.51 & 751.4 & 482.3 & 0.35 \\
Dipole & Off & 99.6  & 52.8  & 0.47 & 435.4 & 161.9 & 0.63 & 462.7 & 213.6 & 0.54\\
       & On  & 271.8 & 108.2 & 0.60 & 831.7 & 286.6 & 0.65 & 859.2 & 340.1 & 0.60\\
\hline 
\end{tabular}
\caption{Time taken~(s) to simulate 10000 $gg\to h^0$ events with $\sqrt{s}=13\,$TeV
  for the angular-ordered and dipole showers.  The time taken to
  simulate three scale variations
  $((\mu_R,\mu_F)/\sqrt{2},(\mu_R,\mu_F),\sqrt{2}(\mu_R,\mu_F))$ by direct
  simulation and reweighting is shown together with the fractional difference
  in times, {\it i.e.}  $(T({\rm direct})-T({\rm reweight}))/T({\rm
    direct})$. Events with only the hard process were simulated as well as
  events with multiple-parton interactions~(MPI) both with and without varying
  the factorization and renormalization scales for the secondary scattering
  processes.  }
\label{tab:time}
\end{table*}

 As can be seen the results of calculating the scale variation using our
 reweighting approach are in excellent agreement with those obtained from
 directly simulating the events with the modified scale choice in both cases.

\begin{figure}[h]
\includegraphics[width=0.5\textwidth]{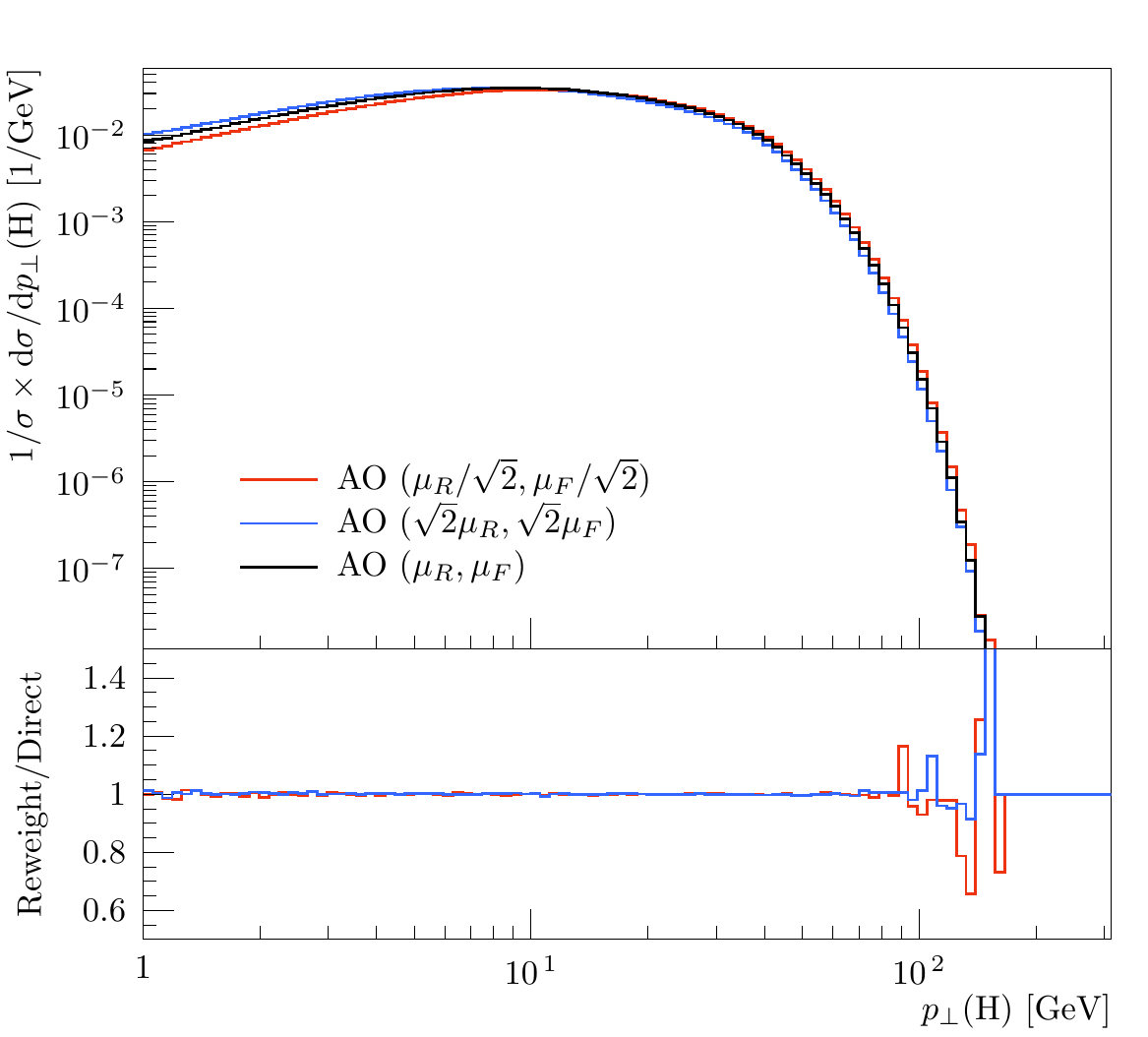}
\caption{The transverse momentum of the Higgs boson in the process $gg\to h^0$
  at the parton level with $\sqrt{s}=13\,$TeV using the angular-ordered parton
  shower.  The upper frame shows the effect of varying the scale while the
  lower shows the ratio of the up and down scale variations calculated using
  reweighting with respect to those obtained running the event generator with
  the relevant scale.}
\label{fig:higgsAO}
\end{figure}

\begin{figure}[h]
\includegraphics[width=0.5\textwidth]{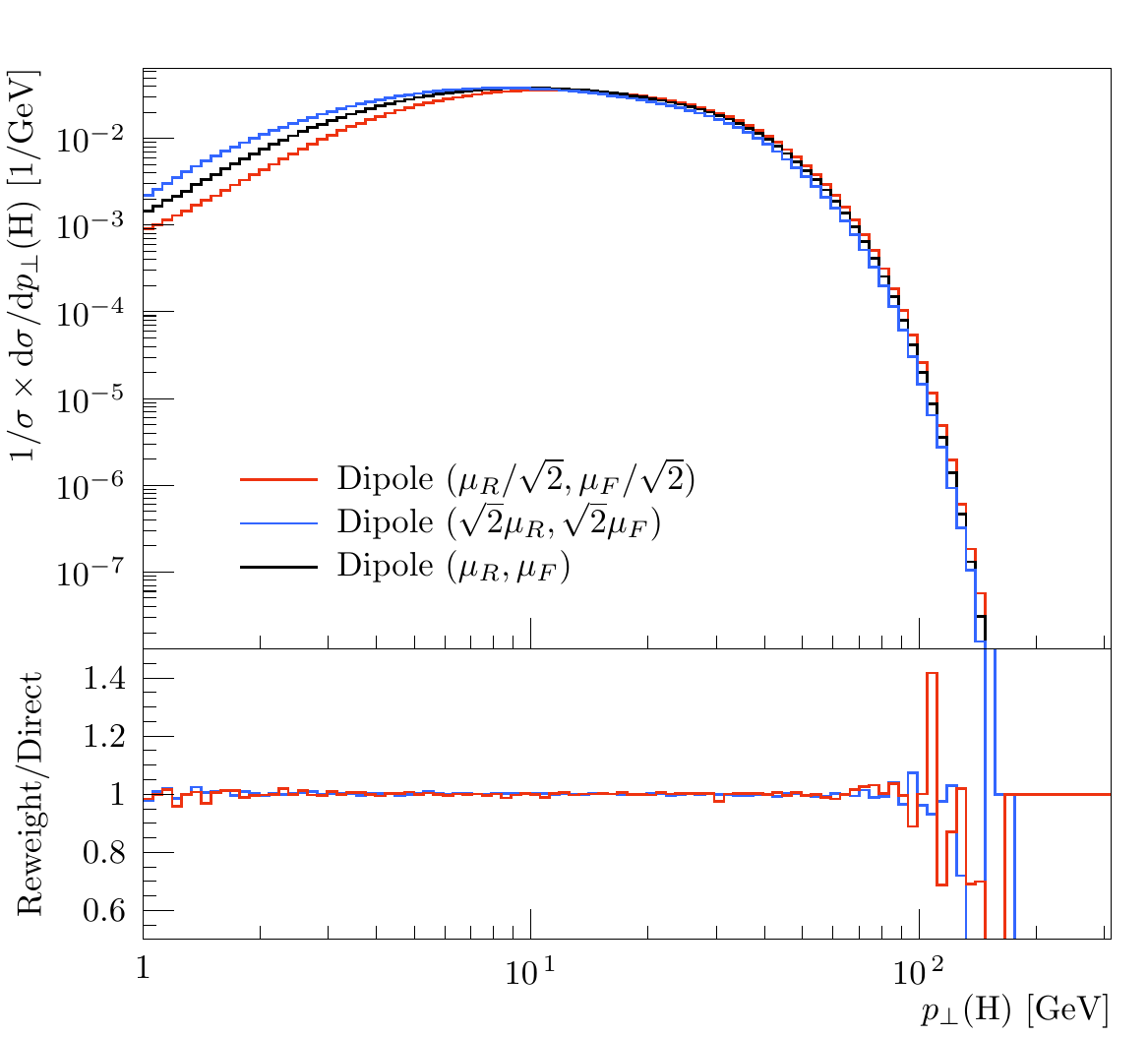}
\caption{The transverse momentum of the Higgs boson in the process $gg\to h^0$
  at the parton level with $\sqrt{s}=13\,$TeV using the dipole parton shower.
  The upper frame shows the effect of varying the scale while the lower shows
  the ratio of the up and down scale variations calculated using reweighting
  with respect to those obtained running the event generator with the relevant
  scale.}
\label{fig:higgsDipole}
\end{figure}

\section{Discussion}

There are two main issues which effect the practicality of using our reweighting
approach to calculate the scale uncertainty in the parton shower:
\begin{enumerate}
\item The time taken to calculate the result of the scale variations using
  reweighting should be less than running the event generator for the
  different scale choices considered. In general this will always be the case
  if the other stages of the event generation, for example the hard process
  evaluation, take significantly longer than the generation of the parton
  shower, or if detector simulation is included. However, for simple processes
  without detector simulation the time taken for the two approaches can be
  comparable, at least for the angular-ordered parton shower,
  Table~\ref{tab:time}.

\item If the weight variation is large then a large number of events will have to be
  simulated in order for the reweighted result to converge on that generated
  by directly simulating the events with an acceptable error. This can be
  particularly problematic if there are regions of phase space which would be
  populated with varied scales which are not filled for the central value, and
  hence have infinite weight.

\end{enumerate}

  The difference in the time taken with the two different shower algorithms is
  due to the different technical implementations of the veto algorithm. The
  dipole shower which uses an adaptive-sampling approach in which only one
  acceptance probability is calculated shows a significant reduction in the
  time taken for the simulation using reweighting because only one additional
  weight needs to be calculated for each variation.
  
  The situation is very different for the angular-ordered parton shower where 
  Eqn.\ref{eqn:vetoprob} is split into a number of different components.
  For example for space-like evolution the splitting kernel is
\begin{equation}
P(q,z) = \frac1q\frac{\alpha_S(z(1-z)q)}{2\pi}P_{\rm AP}(z,q)
\frac{\frac{x}zf(x/z,q)}{xf(x,q)},
\end{equation}
where $q$ is the angular-ordered evolution variable, $x$ is the momentum
fraction of the branching parton and $P_{\rm AP}$ is the Altarelli-Parisi
splitting function. A simple overestimate can be written as
\begin{equation}
R(q,z) = \frac1q\frac{\alpha^{\rm over}_S}{2\pi}P^{\rm over}_{\rm AP}(z) {\rm PDF}^{\rm over}(z),
\end{equation}
where $\alpha^{\rm over}_S$, $P^{\rm over}_{\rm AP}(z)$ and $ {\rm PDF}^{\rm
  over}(z)$ are the overestimates of $\alpha_S(z(1-z)q)$, $P_{\rm AP}(z,q)$
and $\frac{\frac{x}zf(x/z,q)}{xf(x,q)}$, respectively.  The veto is then
separately applied for the weights
\begin{IEEEeqnarray}{LR}
w_1 = \frac{\alpha_S(z(1-z)q)}{\alpha^{\rm over}_S}, \qquad & w_2 =
\frac{P_{\rm AP}(z,q)}{P^{\rm over}_{\rm AP}(z)}, \nonumber
\\ \IEEEeqnarraymulticol{2}{C}{w_3 =
  \frac{\frac{\frac{x}zf(x/z,q)}{xf(x,q)}}{{\rm PDF}^{\rm over}(z)}.}
\end{IEEEeqnarray}
This calculation is organised so that the most time consuming piece, {\it
  i.e.}  the evaluation of $w_3$, is only performed if the emission is accepted
after the tests on $w_{1,2}$.  However, using the reweighting approach all the
weights have to be evaluated for each trial emission, both for the default
scale and any variations, therefore the shower evolution can be slower if
the time taken for the evaluation of these weights is
significant compared to that for the rest of the shower evolution.

This can be seen in Table~\ref{tab:time} where the angular-ordered parton
shower is faster when the scale is varied directly. However, due to the
additional weights which need to be calculated it is
slower than the dipole shower, where the more sophisticated sampling of the
Sudakov form factor means fewer additional weights have to be calculated, when
reweighting is used.

However, when all the necessary parts of the simulation are included even for
the relatively simple hard scattering processes considered here the
reweighting approach is significantly faster and this performance improvement
will only increase when more complicated, and hence time-consuming, processes
are simulated.

Owing to the need to divide out the veto probability in the reweighting
procedure, weight distributions of the reweighted results may broaden
significantly for very efficient sequences of the veto algorithm,
$\epsilon(q,x|k,y) \sim 1$~(see Eqn.\,\ref{eqn:weights}).
While, for the central value, a very efficient
algorithm is desirable it may at the same time force us to use larger
statistics to obtain convergent results for the reweighted
distributions. 

The situation can be improved by explicitly making the veto
algorithm for the central prediction more inefficient than originally designed
by introducing a `detuning' parameter $\lambda > 1$ to increase the proposal
kernel, $R\to \lambda R$. A faster convergence of the reweighted results can
hence be obtained. Despite the increase in the run time, using reweighting is still expected
to be faster than a full simulation of each
variation. Detuning parameters are available for the reweighting mechanisms in
both \hw\ parton shower algorithms.

We show an example of the improvements in weight
distributions in Fig.~\ref{fig:weights}. The negative weight events
appearing for the `down' variation are significantly reduced for a moderate
increase in run time (with $\lambda=4$ taking about twice as long as no
detuning). This increase in run time needs to be compared
with the longer time taken to obtain a similar 
statistical error for the reweighted distributions without detuning
as a  large increase in the
number of simulated events (and hence run time) will be required.
 We note that for the case
of the overestimate kernel being an overestimate for all variations, no
negative weights appear. This is the case, 
for example, when varying the scale in strong coupling, but not the PDFs,
in the angular-ordered parton shower.

\begin{figure}
\includegraphics[width=0.5\textwidth]{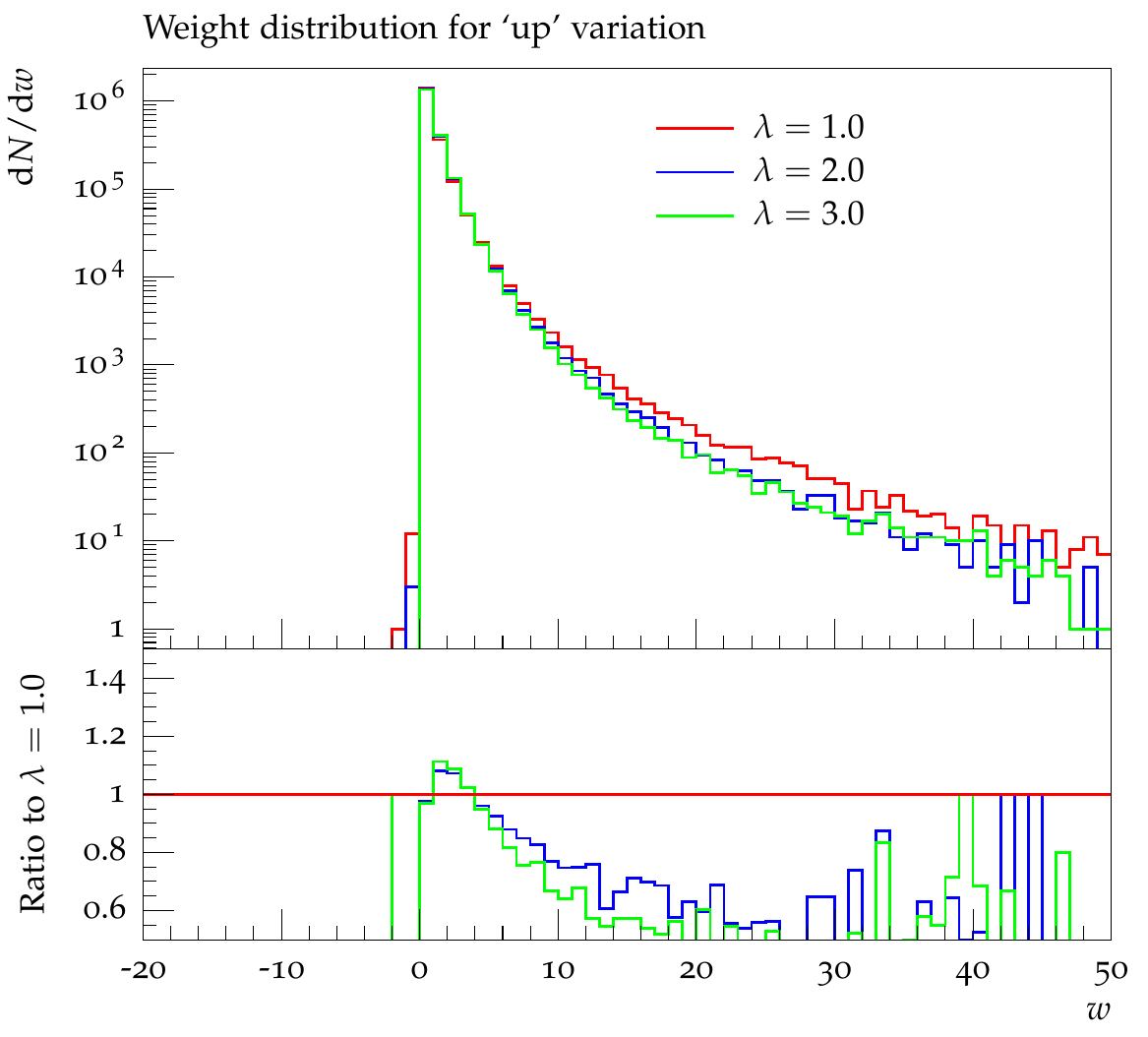}
\includegraphics[width=0.5\textwidth]{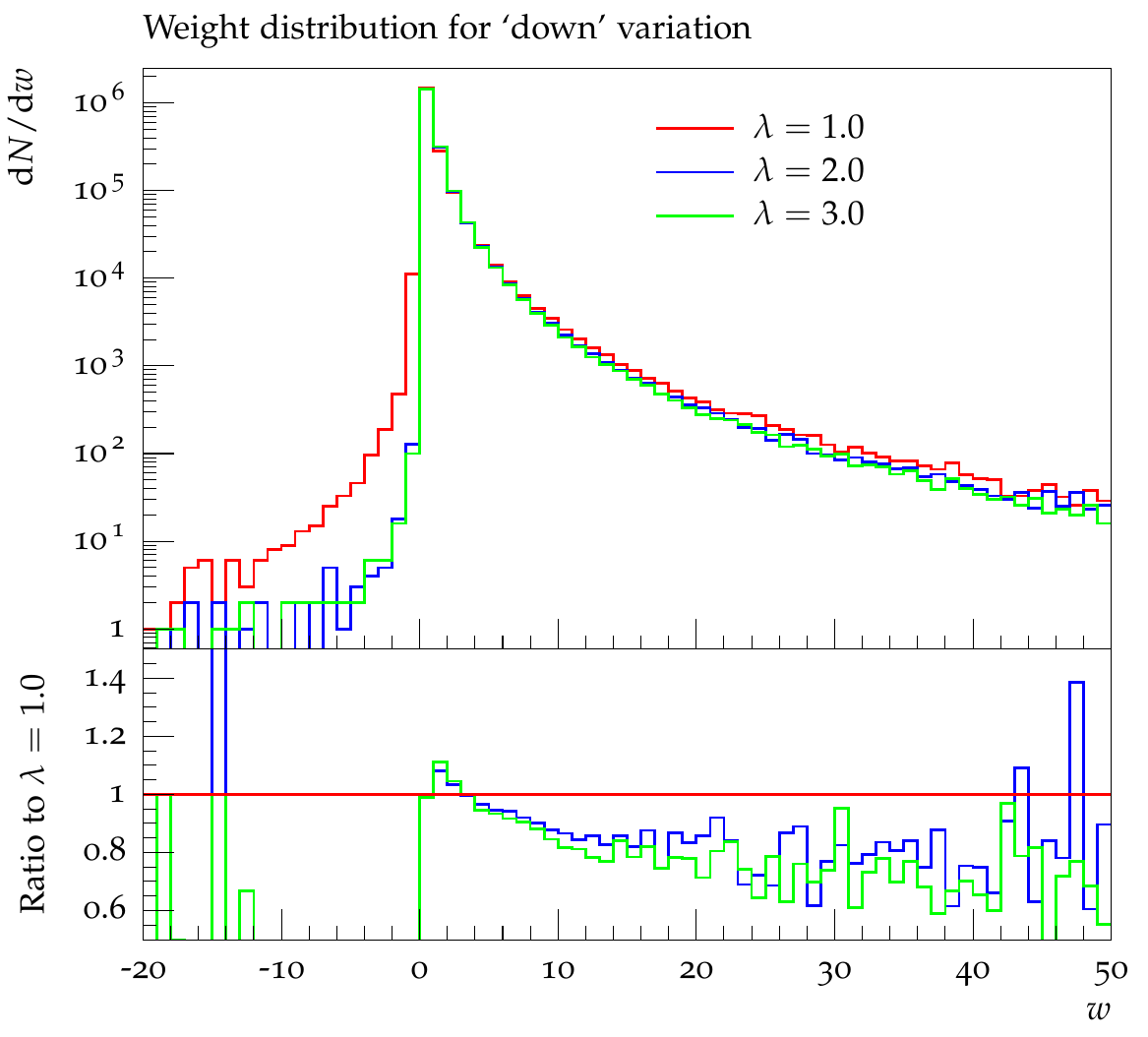}
\caption{Distributions of the weights used to determine the `up' and `down' scale
  variations for the dipole shower in $gg\to h^0$ events at $13\ {\rm TeV}$ for
  different values of the detuning parameter. The improvement by using
  $\lambda > 1$ is clearly observed.\label{fig:weights}}
\end{figure}

\section{Conclusions}

We have presented a new algorithm to allow the inclusion of weights in the
Sudakov veto algorithm.  This new weighted Sudakov veto algorithm allows the
computation of the weights for any variations in splitting kernel used in the
veto algorithm at the same time as the central value is calculated, allowing
efficient computation of the shower uncertainties.

This allows us to assess the uncertainty due to variations of the scales in
the parton shower without resimulating the events for each scale choice of
interest. This is significantly faster and makes the calculation of the scale
uncertainty feasible as the time taken to simulate an individual event can be
time consuming for complicated processes.

This new approach is available in the \hw~(7.0.2) release and will be combined
with the effect of varying the scales in the calculation of the hard process
in a future release.  This technique can be extended to include the effect of
varying the parton distribution functions or changes to the splitting kernel.

\section*{Note Added in Proof}

While this work has been finalized, a similar approach by the \textsf{Pythia}
collaboration \cite{mrennaSkands} came to our attention, and the
\textsf{Sherpa} collaboration has also reported comparable functionality in
\cite{Badger:2016bpw}.

\section*{Acknowledgements}

This work was supported in part by the European Union as part of the FP7
Marie Curie Initial Training Network MCnetITN (PITN-GA-2012-315877). 
It was also supported in part by the 
Institute for Particle Physics Phenomenology under STFC grant ST/G000905/1.

SP acknowledges support by a FP7 Marie Curie Intra European
Fellowship under Grant Agreement PIEF-GA-2013-628739.

\bibliography{reweighting}

\end{document}